\RequirePackage[2020-02-02]{latexrelease}
\documentclass[twocolumn,preprintnumbers,aps,prl,longbibliography]{revtex4}
\usepackage{dcolumn}
\usepackage{bm}
\usepackage{color}
\usepackage{float}
\usepackage[caption = false]{subfig}
\usepackage[final]{graphicx}
\usepackage{amsmath}

\begin{document}

\title{Implementation of Shor's Algorithm with a Single Photon in 32 Dimensions}

\author{Hao-Cheng Weng}
\email [Email address: ]{haocheng.weng@bristol.ac.uk}
\altaffiliation[Present affiliation: ]
{Quantum Engineering Technology Laboratories and H. H. Wills Physics Laboratory, University of Bristol, UK.}
\affiliation{Department of Physics and Center for Quantum Science and Technology, National Tsing Hua University, Hsinchu 30013, Taiwan}

\author{Chih-Sung Chuu}
\email [Email address: ]{cschuu@phys.nthu.edu.tw}
\affiliation{Department of Physics and Center for Quantum Science and Technology, National Tsing Hua University, Hsinchu 30013, Taiwan}

\begin{abstract}

Photonics has been a promising platform for implementing quantum technologies owing to its scalability and robustness. In this Letter, we demonstrate the encoding of information in 32 time bins or dimensions of a single photon. A practical scheme for manipulating the single photon in high dimensions is experimentally realized to implement a compiled version of Shor’s algorithm on a single photon. Our work demonstrates the powerful information processing capacity of a high-dimensional quantum system for complex quantum information tasks.   

\end{abstract}


\maketitle

\section{I. Introduction}

The ability to encode information on single photons has changed how information can be processed or transferred fundamentally, enabling advanced technologies such as quantum communication or quantum computing \cite{Kok2007,OBrien2009,Ralph2010,Ladd2010,Slussarenko2019}. Empowered by this ability, a global network capable of exchanging information between quantum devices is now being envisioned \cite{Wehner2018}. The scalability of such a quantum network will hence be strictly limited by how much information a photon can carry. 

To encode a large quantity of information on a photon, one may harness various degrees of freedom (DOF) of a single photon, including the optical angular momentum, frequency bins, or time bins \cite{Barreiro2005,Babazadeh2017,Wang2018,Erhard2018,Giordani2019,Gao2020,Rambach2021, fickler2012quantum}. Alternatively, the information can be encoded in high dimensions \cite{Joo2007,Reimer2019,Wang2020,Cozzolino2019,Erhard2020, wang2018multidimensional, Chi2022} such as the optical modes of a single physical DOF. It is also proposed that, with the computational space largely expanded by the high-dimensional encoding, powerful information processing can be realized with only ``one" photon in high dimensions \cite{Xiaoqin2019, Juan2011, Kysela2020, kagalwala2017single, abouraddy2012implementing}. Nevertheless, precise control over various optical properties in large dimensions is very challenging at the single-photon level. The experimental implementation of quantum algorithms such as Shor's algorithm using one high-dimensional photon alone has never been achieved.

In this Letter, we explore the information encoding and manipulation of a high-dimensional single photon in a single physical DOF. We demonstrate the encoding of information on a temporally long single photon across 32 time bins or dimensions, which is the largest reported to date for a time-bin-encoded single photon. The high-dimensional quantum state is efficiently prepared by modulating the single photon wave packet in a single shot. By manipulating the high-dimensional single-photon state with a compact interferometry and time-resolved detection, we experimentally realize a compiled version of Shor's algorithm \cite{Vandersypen2001,Lanyon2007,Lu2007,Alberto2009} and the factorization of the integer 15 using a single photon for the first time. Shor's algorithm \cite{Shor1994} is the most sophisticated quantum algorithm with an exponential speed-up on the factorization problem. Our work thus demonstrates not only the astonishing information capacity of a single photon but also the powerful information processing of a single photon in high dimensions. 

\begin{figure}[tbp]
\includegraphics[width=1\columnwidth]{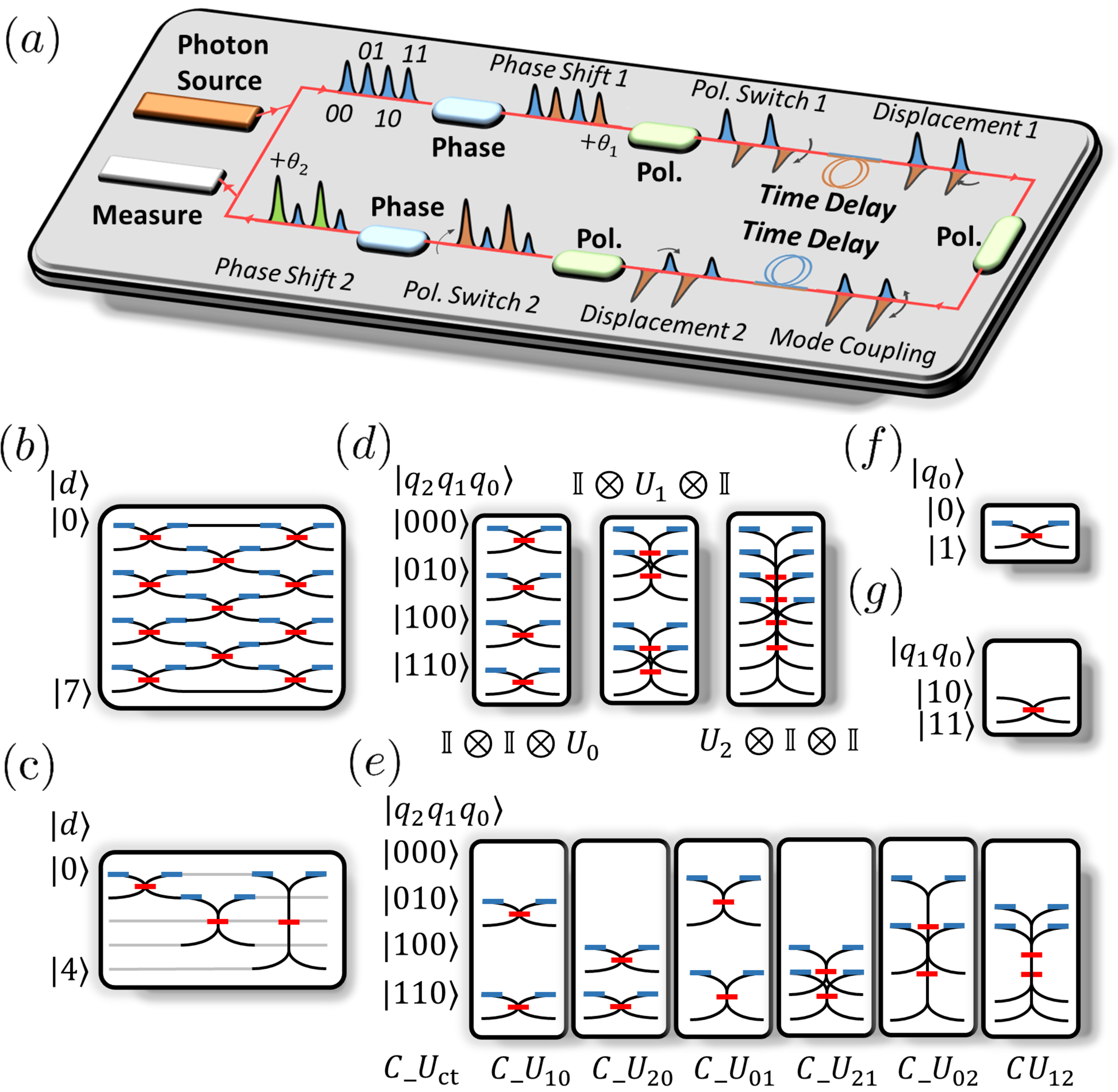}
\caption{(a) Example for manipulating a two-qubit state encoded in 4 time-bin modes. ``Phase" and ``Pol." represent the electro-optic phase and polarization modulators, respectively. The time-bin modes with different phases, which are in one of two orthogonal polarizations, are labeled in different colors. The time delay is implemented in one polarization as indicated by its color. (b) The interferovmetry after several rounds for a constant time delay and nearest neighboring mode coupling prepares arbitrary qudit states. The states $|d\rangle$ are in the \textit{d}-level Hilbert space. (c) The mode coupling with variable time delays. (d) Single-qubit unitary rotations $U_0$, $U_1$, and $U_2$ for a 3-qubit state $|q_2 q_1 q_0\rangle$ encoded on 8 time-bin modes. (e) Realization of the rotational gates with a single-round operation acting on a qubit state. (f) Two-qubit control unitary gates $C\_U_{ct}$ for a two-qubit state encoded in 4 time-bin modes. $c$ and $t$ stand for the control and target qubits, respectively. (g) Realization of the $\textrm{CNOT}_{10}$ gate for a two-qubit state encoded in 4 time-bin modes.}
\label{fig1}
\end{figure}

\section{II. The Quantum Circuit}

The compiled version of Shor's algorithm \cite{Vandersypen2001,Lanyon2007,Lu2007,Alberto2009} is designed to find the prime factors of a specific integer $N$ or, equivalently, the order $r$ of $a\ \textrm{modulo}\ N$ such that $1 \leq r<N$ and $a^r \equiv 1\ \textrm{mod}\ N$ for any coprime integers $a$ and \textit{N}. The algorithm consists of three steps. The \textit{register initialization} prepares the first (argument) register in a equal coherent superposition state,
\begin{align}
\begin{split}
|0\rangle^{\otimes n} |0\rangle^{\otimes m} \rightarrow \frac{1}{\sqrt{2^n}} \sum_{x=0}^{2^n-1}|x\rangle|0\rangle^{\otimes m-1}|1\rangle.
\end{split}
\end{align}
The \textit{modular exponentiation} then produces the highly entangled state,
\begin{align}
\begin{split}
\frac{1}{\sqrt{2^n}} \sum_{x=0}^{2^n-1}|x\rangle |a^x \textrm{mod} N\rangle
\end{split}
\end{align}
by applying the order-finding function $a^x \textrm{mod} N$ to the second (function) register when the argument register is in the state $|x\rangle$. Finally, the \textit{inverse Quantum Fourier Transform} (QFT) is applied to the argument register,
\begin{align}
\begin{split}
\frac{1}{2^n} \sum_{y=0}^{2^n-1}\sum_{x=0}^{2^n-1} e^{2\pi i xy/2^n} |y\rangle |a^x \textrm{mod} N\rangle,
\end{split}
\end{align}
followed by the measurement in the computational basis to find the interference peak at $|y\rangle$ with $y=c2^n/r$. 

To encode the computational basis states on a single photon, we exploit the time-bin modes $d_j$ within its temporal wave packet,
\begin{align}
\begin{split}
\sum_{j=1}^{2^n}c_j|d_j\rangle =&\ c_1 |00...0\rangle+c_2 |00...1\rangle+... \\
&+c_{2^n-1}|11...0\rangle+c_{2^n}|11...1\rangle, 
\label{eq1}
\end{split}
\end{align}
where $2^n$ is the number of dimensions and $c_j$ is the probability amplitude of the \textit{j}-th time-bin mode. To manipulate the time-bin modes, the single photon is prepared in an equal superposition state before entering a loop structure of programmable phase shifters and time-bin mode couplers (Fig. 1). The phase shifter implements the phase modulation $e^{i\phi}$ for individual time-bin mode. The time-bin mode coupler, composed of sequential operations of full polarization rotation (pol. switch), time-bin displacement (time delay), and partial polarization rotation, couples different time-bin modes $d_n$ and $d_m$ \cite{Humphreys2013} as follows,
\begin{gather}
\begin{pmatrix}
  d_n' \\ 
  d_m'
\end{pmatrix}=
\begin{pmatrix}
  \sqrt{1-C} & \sqrt{C}\\ 
  -\sqrt{C} & \sqrt{1-C}
\end{pmatrix} 
\begin{pmatrix}
  d_n \\ 
  d_m
\end{pmatrix},
\end{gather}
where $C$ is the coupling strength. Consider a constant time delay such that the mode coupling only exists between the nearest neighboring time-bin modes. The repeated operations structured in Fig. 1(b), where the blue (red) lines represents the phase shifters (mode couplers), generates arbitrary state in the form of Eq. (4) \cite{Motes2014,Humphreys2013,He2017}. If more efficient coupling between the ``far-away" time-bin modes is needed, tunable time-bin displacement can be exploited as shown in Fig. 1(c). 

The elementary gates can be realized with the phase shifters and mode couplers. Consider a three-qubit state encoded in 8 time-bin modes. The single-qubit unitary operations $U_n$ is realized in Fig. 1(d). Targeted on different qubit, the operation is implemented by the mode coupling with different time-bin displacements. For arbitrary single-qubit gates, which can be decomposed into the single-qubit rotations $e^{i\theta \hat{\sigma_y}}$ and $e^{i\theta \hat{\sigma_z}}$, the programmable phase shift and mode coupling operations in Fig. 1(e) are employed. For general two-qubit control unitary operations, the realization for a three-qubit state encoded in 8 time-bin modes is shown in Fig. 1(f). The time-bin displacement is again dependent on the target qubit. As an example, Fig. 1(g) shows how $\textrm{CNOT}_{10}$ for a two-qubit state encoded in 4 time-bin modes can be implemented by swapping the states $|10\rangle$ and $|11\rangle$.

\begin{figure}[tbp]
\includegraphics[width=1\columnwidth]{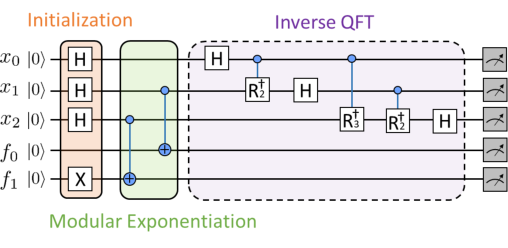}
\caption{Quantum circuit for implementing the compiled version of Shor's algorithm that factorizes 15. $x_0$, $x_1$, and $x_2$ are the argument registers while $f_0$ and $f_1$ are the function registers. $R_n$ is the phase shift $e^{2\pi i/2^n}$.}
\label{fig2}
\end{figure}

Figure 2 shows the quantum circuit used for our implementation of Shor's algorithm with $N=15$ and $a=2$. The state $|f_1 f_0 x_2 x_1 x_0 \rangle$ is initialized to
\begin{align}
\begin{split}
    ( &|10000\rangle+|10001\rangle+|10010\rangle+|10011\rangle \\
		&+|10100\rangle+|10101\rangle+|10110\rangle+|10111\rangle)/{\sqrt{8}}, \\
\label{eq2}
\end{split}
\end{align}
after the register initialization and evolves into
\begin{align}
\begin{split}
    ( &|00100\rangle+|00101\rangle+|01110\rangle+|01111\rangle \\
		&+|10000\rangle+|10001\rangle+|11010\rangle+|11011\rangle)/\sqrt{8}
\end{split}
\end{align}
after the modular exponentiation. Since $r=4=2^m$ for natural numbers $m$, the inverse QFT can be carried out by the classical processing \cite{Griffiths1996,Lanyon2007,Lu2007,Alberto2009}.

\section{III. Experimental Setup}

The experimental setup for implementing the Shor's algorithm is illustrated in Fig. 3. Type-II phase-matched bipthons are generated by the doubly resonant parametric down-conversion in a monolithic PPKTP crystal \cite{Chuu2012,Wu2017}. By pumping the crystal with a cw laser at 775 nm, the detection of one photon heralds a single photon at 1550 nm. The wave packet of the heralded single photon, measured by the time-resolved coincidence counting and shown in Fig. 5(a), has a $1/e^2$ coherence time of 148~ns with a bandwidth of 2.7 MHz \cite{Chuu2011} and pair generation rate of $4.2 \times 10^5$ s$^{-1}$mW$^{-1}$. The temporally long wave packet not only enables the manipulation of the time-bin modes by electro-optic modulators (2 dB insertion loss) \cite{Kolchin2008,Feng2017,Wu2019,Chinnarasu2020,Cheng2020,Chuu2021,Kao2023} but also allows us to initialize the state in Eq. (6) by the electro-optic amplitude modulation in just one shot. This shows the advantage of using narrowband single photons to generate a large number of time bins as compared to multi-passing the loop structure \cite{He2017, chen2018observation, schreiber2010photons} in terms of the setup complexity and scalability. It is also worth noting that, to prepare the initial state in Eq. (6), no phase modulation is needed as all the time bins are in phase.

\begin{figure}[tbp]
\includegraphics[width=1\columnwidth]{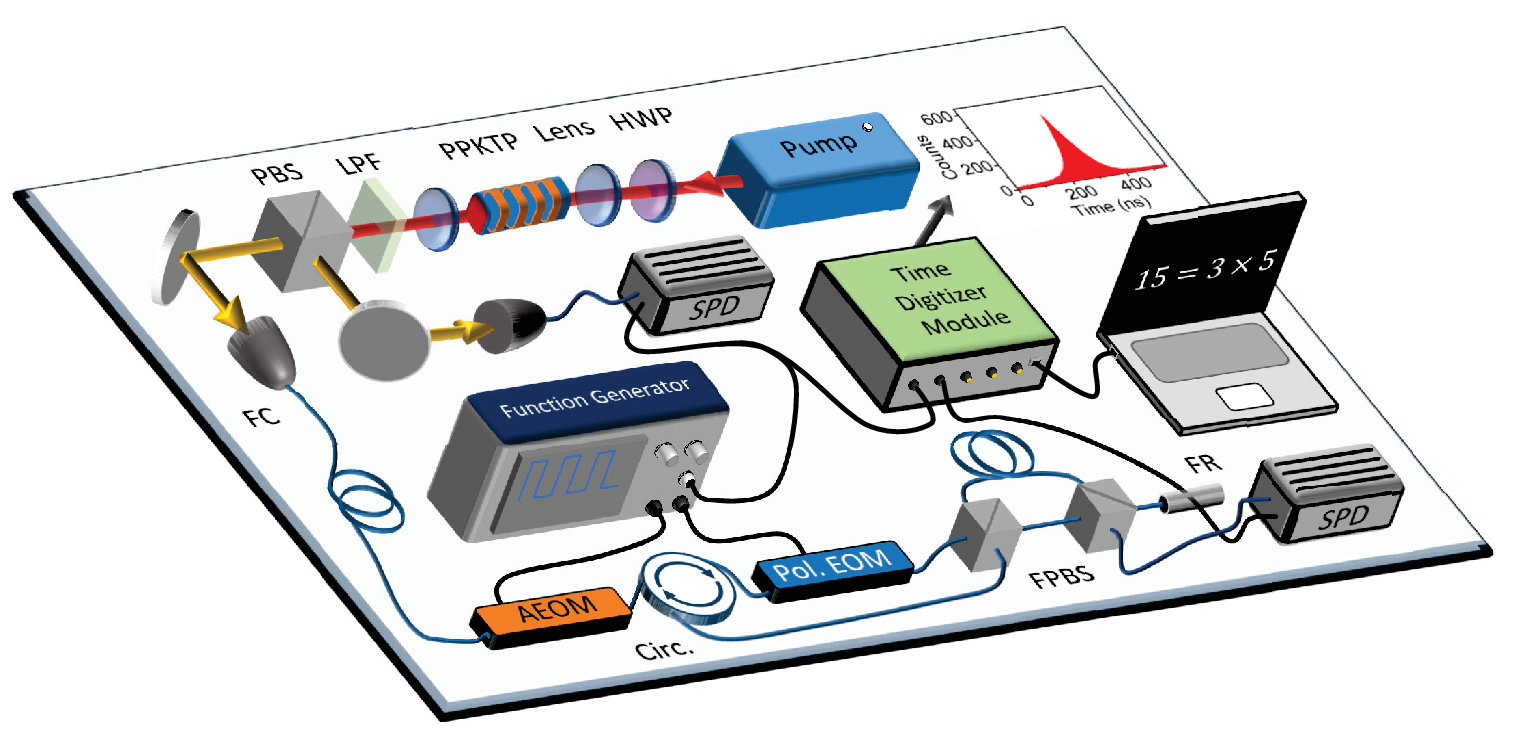}
\caption{Experimental setup of the implementation of Shor's algorithm in a single photon. The monolithic PPKTP crystal is pumped by a frequency-locked cw laser at 775 nm to generate type-II phase-matched biphotons. The half-wave plate (HWP) aligns the pump polarization to the crystal. A pair of lenses is used to focus the pump beam into the crystal. The long-pass filter (LPF) is placed to filter out the pump. After the polarizing beam splitter (PBS) splits the generated biphoton, the heralded 1550 nm single photons are sent into the fiber-based interferometry for implementing the Shor's algorithm, where AEOM, Pol. EOM, and SPD stand for the electro-optic amplitude modulator, electro-optic polarization modulator (switch), and single-photon detector (15\% quantum efficiency, 150 ps timing resolution), respectively. A 200-ns optical delay is used for delaying the single photons.}
\label{fig3}
\end{figure}

To simplify the implementation of the gate operations and minimize the optical loss associated with the fiber-based devices, optical circulator and fiber optic retroreflector are used to enable the multi-passing of the optical loop in a more compact architecture. The two CNOT gates in the modular exponentiation are implemented by an electro-optic polarization switch (3.5 dB insertion loss) and optical delays (fibers). While the first CNOT gate double-passes the optical delay by using two fiber-based polarizing beam splitters, the second CNOT gate passes the optical delay once. After the modular exponentiation, the outcome is analyzed by a time digitizer (100 ps time resolution) and processed by a computer to reveal the result of the factorization.

\section{IV. Results}

The single-qubit and two-qubit gates, which compose an universal gate set \cite{nielsen2002}, are carefully characterized in Fig. 4. The measured single-qubit rotations $e^{i\phi \hat{\sigma_y}}$ and $e^{i\phi \hat{\sigma_z}}$ on a Bloch sphere are shown in Fig. 4(a) and Fig.4(b), respectively. These single-qubit rotations are used not only for implementing arbitrary single-qubit gates but also for calibrating the modulation signals. The measured truth table of the $\textrm{CNOT}_{10}$ gate is shown in Fig. 4(c), which is in good agreement with the theoretical truth table in Fig. 4(d).

\begin{figure}[tbp]
\includegraphics[width=1\columnwidth]{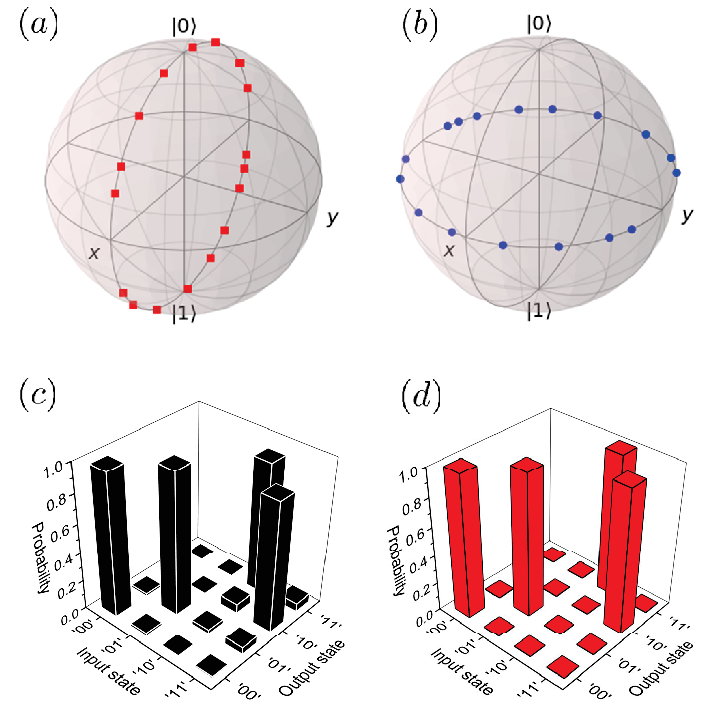}
\caption{Characterization of the single-qubit rotations and CNOT gate. (a) Single-qubit rotation $e^{i\phi \hat{\sigma_y}}$. (b) Single-qubit rotation $e^{i\phi \hat{\sigma_z}}$. The data points are not evenly spaced because the phase shifts are not linearly dependent on the voltages, which are linearly varied, applied to the electro-optic phase modulators. (c) The measured truth table of the $\textrm{CNOT}_{10}$ gate. (d) The theoretical truth table of the $\textrm{CNOT}_{10}$ gate.}
\label{fig4}
\end{figure}

The experimental demonstration of Shor's algorithm with a single photon is summarized in Fig. 5 and 6. As shown in Fig. 5(b) to 5(d), the initial time-bin modes as well as the output state probabilities after the first CNOT gate and modular exponentiation exhibit excellent visibility with uneven amplitudes. The uneven amplitudes are due to the double-exponential single-photon wave packet [Fig. 5(a)] and the different losses experienced by different time-bin modes. Nevertheless, Fig. 6 shows that the outcome of the argument registers $x_2 x_1 x_0$ after the inverse QFT is in good agreement with the theoretical prediction assuming even amplitudes for the initial time-bin modes. This surprising result is a consequence of the Hadamard operations in the inverse QFT, which equally distribute the interference peaks at $000$, $010$, $100$, and $110$. These peaks correspond to $x = c 2^3 / r = 0, 2, 4, 6$, respectively, for integer $c$. The case $x=0$ is an inherent failure of Shor's algorithm while the case $x=4$ gives the trivial result of the factors $1$ and $15$. The cases $x=2$ and $x=6$ result in finding the order (periodicity) $r=4$ with the factors $\textrm{gcd}(a^{r/2}\pm 1,N)=3$ and $5$.

\begin{figure}[tbp]
\includegraphics[width=1\columnwidth]{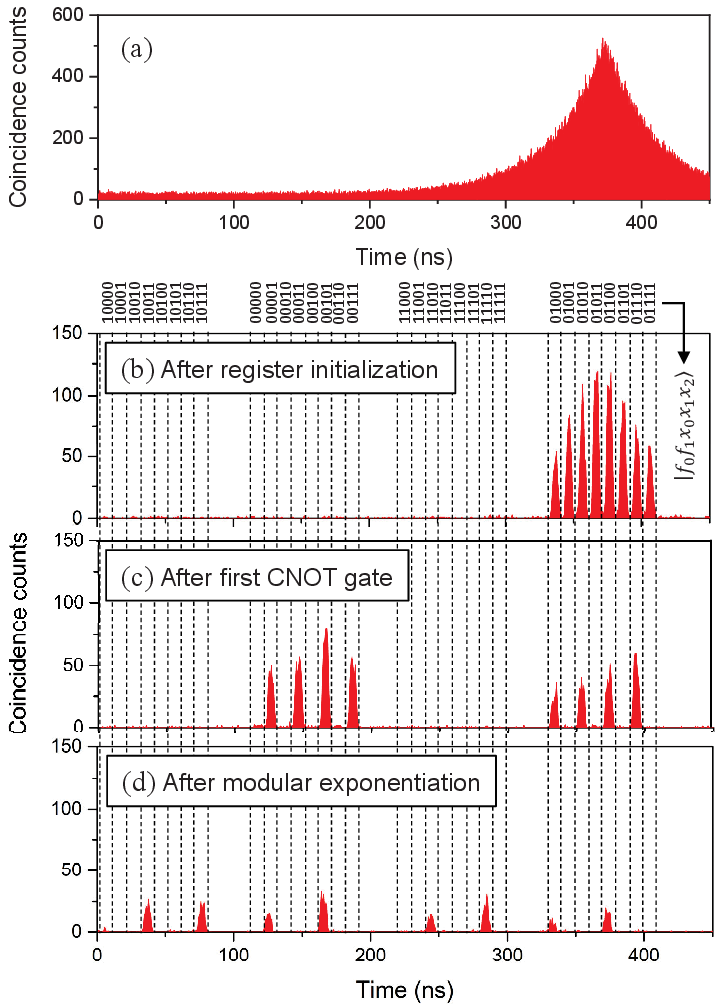}
\caption{(a) The wave packet of the narrowband single photons has a double-exponential waveform. The output state probabilities after the register initialization, first CNOT gate, and modular exponentiation are shown in (b), (c), and (d), respectively. The dashed lines indicate the boundaries of the time-bin modes, which correspond to different computational basis states $|f_0 f_1 x_0 x_1 x_2 \rangle$ (the order of the qubits are altered for convenience). For the result in (d), a coincidence count rate of 10 counts/s is observed and the statistics is accumulated for 30 minutes.}
\label{fig5}
\end{figure}

\begin{figure}[tbp]
\includegraphics[width=1\columnwidth]{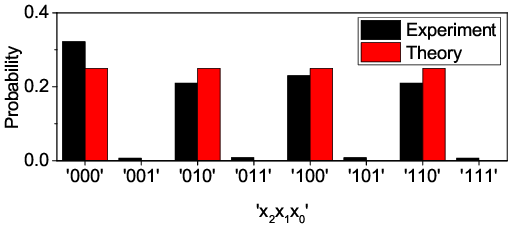}
\caption{Experimental order finding after the inverse QFT of the Shor's algorithm. The theory considers initial time-bin modes with even amplitudes.}
\label{fig6}
\end{figure}

\section{V. Conclusion}

In summary, we have implemented Shor's algorithm with a single photon by encoding and manipulating 32 time-bin modes on the single-photon wave packet. The ability of performing such a complex quantum computing task manifests the powerful information processing capacity of a single photon in high dimensions. Considering commercially available electro-optic modulators with a 40 GHz bandwidth (EOSPACE, Inc.), encoding more than 5000 time-bin modes on  temporally long single photons is possible. The manipulation and noise with the high-dimensional states are usually more troublesome compared to the qubits. Nevertheless, our work shows that the high-dimensional time-bin states of a temporally long single photon can be prepared in a single shot and manipulated by a compact programmable fiber loop. The high-dimensional states may also be manipulated by the high-dimensional quantum gates \cite{Babazadeh2017, Gao2020} with the single-qudit interferometry replaced by a multi-qudit interferometry \cite{He2017}, in which the use of multiple photons provides the scalability. The reduction of the number of single-photon sources and detectors will be beneficial for achieving a higher ratio of the coincidence to accidental counts \cite{wang2021integrated, christ2012limits,Wang2012,Chi2022}. It has also been shown that the high-dimensional states have a higher resistance to quantum channel noise \cite{Cerf2002,Islam2017,Bouchard2017}. The time-bin-encoded states of temporally long single photons are thus promising for implementing high-dimensional quantum computing.

\section{ACKNOWLEDGMENTS}

The authors would like to thank D.-S. Chuu and C.-W. Yang for the insightful discussion. This work was supported by the National Science and Technology Council, Taiwan (110-2112-M-007-021-MY3, 111-2119-M-007-007, and 112-2119-M-007-007).

\bibliography{Paper2}

\end{document}